\documentclass{aa501}

\usepackage{graphicx}
\usepackage{natbib}
\bibpunct{(}{)}{;}{a}{}{,}

\newcommand{\Mwd}{\mbox{$M_\mathrm{wd}$}}
\newcommand{\Teff}{\mbox{$T_\mathrm{eff}$}}
\newcommand{\Prot}{\mbox{${P_\mathrm{rot}}$}}
\newcommand{\Ha}{\mbox{${\mathrm H\alpha}$}}
\newcommand{\Hb}{\mbox{${\mathrm H\beta}$}}
\newcommand{\Hg}{\mbox{${\mathrm H\gamma}$}}
\begin{document}

\title{Magnetic white dwarfs in the Early Data Release of the Sloan Digital Sky Survey}

\author{B.T. G\"ansicke\inst{1} \and
        F. Euchner\inst{2} \and
        S. Jordan\inst{3}}
\offprints{B. G\"ansicke, e-mail: btg@astro.soton.ac.uk}

\institute{
Department of Physics and Astronomy, University of Southampton,
  Southampton SO17 1BJ, UK
\and
  Universit\"ats-Sternwarte, Geismarlandstr. 11, 37083 G\"ottingen, 
  Germany
\and
 Institut f\"ur Astronomie und Astrophysik, Universit\"at
 T\"ubingen, Sand 1, 72076 T\"ubingen, Germany
}

\date{Received \underline{\hskip2cm} ; accepted \underline{\hskip2cm} }

\abstract{We have identified 7 new magnetic DA white dwarfs in the
Early Data Release of the Sloan Digital Sky Survey. Our selection
strategy has also recovered all the previously known magnetic white
dwarfs contained in the SDSS EDR, KUV\,03292+0035 and
HE\,0330--0002. Analysing the SDSS fibre spectroscopy of the magnetic
DA white dwarfs with our state-of-the-art model spectra, we find
dipole field strengths $1.5\,\mbox{MG}\le B_\mathrm{d}\le 63$\,MG and
effective temperatures $8500\le\Teff\le 39\,000$\,K. As a conservative
estimate, we expect that the complete SDSS will increase the number of
known magnetic white dwarfs by a factor 3.
}

\maketitle

\section{Introduction}
The population of magnetic white dwarfs spans an enormous parameter
space in magnetic field strength $B$, effective temperature \Teff,
rotational period \Prot, atmospheric abundances, and mass \Mwd\ --
with the number of accurate measurements per parameter dimension
decreasing in this sequence.  
Despite intense spectroscopic and polarimetric surveys carried out
over the last two decades \citep[e.g.][]{schmidt+smith95-1,
putney95-1, putney97-1, hagenetal87-1, reimersetal94-1,
reimersetal96-1, reimersetal98-1}, only $\sim65$ magnetic white dwarfs
are known at present \citep{jordan01-1,wickramasinghe+ferrario00-1}. The small
size of this sample seriously hampers the progress of our understanding of
the origin of the strong magnetic fields found in a small fraction
(few \%) of all white dwarfs, as well as of the evolution of these
exotic stars.

The Sloan Digital Sky Survey (SDSS), the largest spectroscopic survey
carried out to date, samples a great variety of galactic and
extragalactic objects at high galactic latitudes. Due to the
partial overlap in colour space between white dwarfs and quasars, it
can be expected that the SDSS will result in the identification of a
large number of white dwarfs and, hence, in a significantly increased
sample of known magnetic white dwarfs. Here we describe the sample of
magnetic white dwarfs identified in the Early  Data Release of the SDSS.

% ---------- Table 1 -------------------------------------------------------
\begin{table*}[t]
\caption[]{\label{t-sdssobs} Confirmed and candidate magnetic white
dwarfs from the SDSS EDR. The equinox/epoch 2000 coordinates are coded
in the object designation. The objects are classified in the SDSS EDR
as stellar (S) or unknown (U). The spectra are uniquely identified in
the SDSS database by the combination of the Plate ID, the modified
Julian date, and the fibre ID of the observation. The magnitudes
listed here are from the associated imaging data.}
\begin{flushleft}
\setlength{\tabcolsep}{1.2ex}
\begin{tabular}{llllllccccc}
\hline\noalign{\smallskip}
MWD         & 
\multicolumn{2}{c}{Class. ~~~ Sample} &
Spectrum ID & 
Obs. date   &
Exp.        &
$u^*$ & $g^*$ & $r^*$ & $i^*$ & $z^*$ \\
\hline\noalign{\smallskip}
%----------------------------------------------------
SDSS\,J030407.40--002541.7 & DA & S & 411-51817-172 & 
2000-09-30 10:37 & 2700\,s &
18.06 & 17.77 & 17.95 & 18.09 & 18.37 \\ %S
%----------------------------------------------------
SDSS\,J033145.69+004517.0$^*$ & DA & S & 415-51810-370 &
2000-09-23 08:34 & 3600\,s & 
17.31 & 17.23 & 17.49 & 17.74 & 18.00 \\ %S
%----------------------------------------------------
SDSS\,J033320.37+000720.7$^+$ & DB?\, & U &415-51810-492 &
2000-09-23 08:34 & 3600\,s &
17.02 & 16.52 & 16.41 & 16.34 & 16.48 \\ %E
%----------------------------------------------------
SDSS\,J034511.11+003444.3 & DA & S & 416-51811-590 &
2009-09-24 09:35 & 3600\,s &
19.11 & 18.63 & 18.52 & 18.49 & 18.50 \\ %S
%----------------------------------------------------
SDSS\,J121635.37--002656.2 & DA & U & 288-52000-276 &
2001-04-01 06:24 & 3602\,s & 
19.85 & 19.57 & 19.83 & 20.05 & 20.11 \\ %E
%----------------------------------------------------
SDSS\,J122209.44+001534.0 & DA & U & 289-51990-349 &
2001-03-22 06:25 & 3604\,s &
20.51 & 20.23 & 20.47 & 20.65 & 21.04 \\ %E
%----------------------------------------------------
SDSS\,J172045.37+561214.9 & DA & U & 367-51997-461 &
2001-03-22 11:51 & 6302\,s &
20.00 & 20.11 & 20.49 & 20.76 & 21.29 \\ %E
%----------------------------------------------------
SDSS\,J172329.14+540755.8 & DA & S & 359-51821-415 &
2000-10-03 03:59 & 4500\,s &
19.14 & 18.81 & 18.90 & 19.04 & 19.30 \\ %S
%----------------------------------------------------
SDSS\,J232248.22+003900.9 & DA & S & 383-51818-421 &
2000-09-07 08:10 & 3600\,s &
18.91 & 19.02 & 19.31 & 19.62 & 19.82 \\ %S
%----------------------------------------------------
SDSS\,J232337.55--004628.2 & DA? & S & 383-51818-215 &
2000-10-01 04:29 & 3600\,s &
17.88 & 18.00 & 18.27 & 18.52 & 18.78 \\ %S
%----------------------------------------------------
\noalign{\smallskip}\hline
\multicolumn{9}{l}{$^*$\,=\,KUV\,03292+0035~~~$^+$\,=\,HE\,0330--0002} \\
\end{tabular}
\end{flushleft}
\end{table*}
% -------------------------------------------------------------------------

\section{Sloan Digital Sky Survey observations}
The Sloan Digital Sky Survey is an ambitious project which will
provide deep CCD imaging of $\sim10\,000\,\deg^2$ of the north
Galactic cap in five optical bands, as well as low-resolution
spectroscopy for $\sim10^6$ astronomical objects selected on the base
of their colours derived from the imaging data. Both, imaging and
spectroscopy are carried out with the purpose-built SDSS 2.5\,m
telescope located at Apache Point Observatory.  The SDSS follow-up
spectroscopy is obtained using two fibre-fed spectrographs that allow
to observe more than 600 objects simultaneously. The spectrographs
cover 3800--9200\,\AA\ at a spectral resolution of $R\simeq1800$. The
reduction of the spectral data is performed by an automated software
pipeline. For a technical description of the survey, see
\citet{yorketal00-1}. Even though the main purpose of this project is
to produce the most comprehensive galaxy and quasar redshift survey to
date (the majority of the spectrograph fibres are allocated to
galaxy/quasar candidates) it delivers a vast amount of data for
detailed studies of the galactic stellar population. A few recent
examples of such studies are the discoveries of a very cool white
dwarf \citep{harrisetal01-2}, of a sample of cataclysmic variables
\citep{szkodyetal02-2}, and of a sample of carbon stars
\citep{margonetal02-1}.

The Early Data Release (EDR) of the SDSS \citep{stoughtonetal02-1}
contains imaging data for 462\,$\deg^2$ and a total of 54008 fibre
spectra obtained in that region. 

% ---------- Fig. 1 -------------------------------------------------------
\begin{figure}
\includegraphics[width=8.8cm]{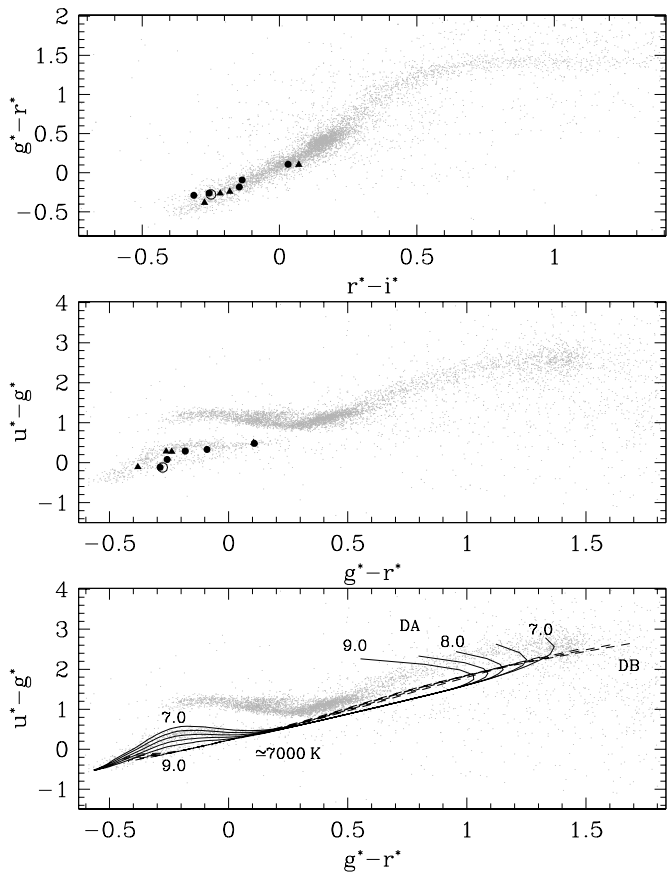}
\caption[]{\label{f-colours} SDSS colour-colour diagrams of the
stellar component of the EDR.  The magnetic white dwarf
candidates selected from the ``stellar'' and ``non-stellar'' fraction
of the EDR sample are plotted as filled circles and filled triangles,
respectively (top and middle panel). Hot stars --~mainly white dwarfs
and subdwarfs~-- are clearly separated from the bulk of the main
sequence stars in the lower left quadrant of the $u^*-g^*/g^*-r^*$
diagram. The bottom panel shows DA (solid lines) and DB (dashed lines)
cooling tracks of non-magnetic white dwarfs for surface gravities
$\log g=7.0$ to 9.0, computed from  model spectra
(P. Bergeron, private communication).  The DA cooling tracks range
from $\Teff=100\,000$\,K (left) to 1500\,K (right) and cross each
other in the gap between the hot stars and the main sequence at
$\Teff\simeq7000$\,K. The DB cooling tracks range from
$\Teff=30\,000$\,K (left) to 4000\,K (right).
The magnetic white dwarf candidates discussed in this paper are well
confined within the hot star population.}
\end{figure}
% -------------------------------------------------------------------------

\section{Selection strategy}
We have used the SDSS Science Archive Query Tool (sdssQT) to select
all entries from the EDR spectroscopic data base that were classified
as ``star'' by the processing pipeline (\texttt{specClass\,=\,1} or
\texttt{6}). A total of 7072 spectra were
downloaded\footnote{http://sdssdp7.fnal.gov/cgi-bin/das/spectra} as
FITS files. All spectra were visually inspected for peculiar
absorption structures, especially for broad absorption troughs near
\Ha\ and \Hb\ as possible indicators for Zeeman splitting in a
magnetic field of a few to a few tens MG. Discarding a number of
spectra with very poor signal-to-noise ratio, six objects were
selected (those marked ``S'' in Table\,\ref{t-sdssobs}). The details
of the observations are listed in Table\,\ref{t-sdssobs}. A
cross-check with the Simbad database revealed that
SDSS\,J033145.69+004517.0 is a known magnetic DA white dwarf,
discovered by \citet{wegneretal87-2} during follow-up observations of
the Kiso survey for UV-excess objects.

Figure\,\ref{f-colours} shows colour-colour diagrams for all 7072
``stellar'' objects from the EDR spectroscopic database. It is
apparent that all six magnetic white dwarfs candidates fall well
within the ``hot star'' quadrant of the $u^*-g^*$ vs. $g^*-r^*$
diagram~--~the vast majority of objects in this colour space are white
dwarfs and subdwarfs. Based on the colour space confinement of the
objects selected  among the ``stellar'' EDR fraction, we extended our
search for magnetic white dwarfs to the spectra classified as
``non-stellar'' by the SDSS processing pipeline
(\texttt{specClass\,=\,0, 2, 3} or \texttt{4}). We downloaded and
visually inspected 3224 spectra whose associated imaging data
satisfies $u^*-g^*<0.6$ and $g^*-i^*<0.5$, and selected four
additional magnetic white dwarf candidates (Table\,1), all of which
were classified as ``unknown'' (\texttt{specClass\,=\,0}).  Cross
correlation with Simbad showed that SDSS\,J033320.37+000720.7 is a
known magnetic white dwarf discovered in the Hamburg/ESO Survey,
probably with a helium-rich atmosphere \citep{reimersetal98-1,
schmidtetal01-2}.

The SDSS fibre spectra of the objects listed in Table\,\ref{t-sdssobs}
are displayed in Figs.\,\ref{f-mwdspectra1} to \ref{f-mwdspectra3}.
We have checked the $u^*g^*r^*i^*z^*$ magnitudes derived from the EDR
spectrum and found the spectroscopic magnitudes to be $\la0.2$\,mag
fainter than the photometric magnitudes for all but the faintest objects,
which is in good agreement with the accuracy of the absolute flux
calibration of the SDSS spectroscopy quoted by
\citet{stoughtonetal02-1}. Monochromatic fluxes at the effective
wavelengths of the $g^*$, $r^*$, and $i^*$ filters derived from the
associated SDSS imaging data are plotted along with the fibre spectra.

\section{Analysis}

We have used the code of \citet{euchneretal02-1} to confirm the
magnetic white dwarf nature of the candidates selected from the SDSS
EDR, and to derive their fundamental properties. In brief, this code
computes flux and circular polarisation spectra for a magnetic white
dwarf with an almost arbitrary field topology by adding up
appropriately weighted model spectra for a large number of surface
elements. To save time, the code makes use of a library of several
10\,000 synthetic spectra calculated with the stellar atmosphere
program developed by Jordan between 1988 and 2002, covering a wide
range in magnetic field strength $B$, effective temperature \Teff\,
and angle between the line of sight and the direction of the magnetic
field at the stellar surface, $\psi$.  Limb darkening is accounted for
by a simple linear scaling law.  The magnetic field configuration of
the white dwarf is described by a multipole expansion of the scalar
magnetic potential. The individual multipole components may be
independently oriented with respect to the rotation axis of the white
dwarf and offset with respect to its centre, allowing for rather
complex surface field topologies. Additional parameters are the white
dwarf effective temperature and the inclination of either the rotation
or the magnetic axis with respect to the line of sight. Observed
spectra can be fitted using an evolutionary strategy with a
least-squares quality function.

For the analysis of the SDSS fibre spectra presented here, we assumed
the simplest field topology, a centred dipole, resulting in four free
parameters: the magnetic dipole field strength $B_\mathrm{d}$, the
effective temperature \Teff, the inclination of the magnetic axis
against the line of sight, $i$, and the flux normalisation.

We were not able to obtain in all cases a fit that consistently
satisfies both the continuum flux distribution and the strength of the
Zeeman absorption features. A similar discrepancy between the strength
of the Balmer lines and the optical continuum slope has been reported
by \citet{achilleosetal91-1} for the low-field white dwarfs WD\,0159--032
($B_\mathrm{d}=6$\,MG, $\Teff=26\,000$\,K) and, to a lesser degree,
for WD\,0307--428 ($B_\mathrm{d}=10$\,MG, $\Teff=25000$\,K).

For \textit{non-magnetic} white dwarfs, the effective temperature can
be derived very reliably from fitting the Balmer line profiles
\citep[e.g.][]{bergeronetal92-1, finleyetal97-1}\footnote{Generally,
in both magnetic and non-magnetic white dwarfs with \Teff\ exceeding
$20\,000$\,K the slope of the continuum is becoming insensitive to the
temperature as it gradually approaches a Rayleigh-Jeans
distribution.}.  However, in the case of magnetic white dwarfs, is is
presently not yet possible to compute detailed line profiles because
there exists no theory that consistently describes Stark broadening in
the presence of substantial magnetic fields \citep[see, e.g.,
][]{jordan92-1}, and, as a result, the computed equivalent widths of
the individual Zeeman components are subject to systematic
uncertainties. Consequently, effective temperatures which are derived
from fitting the Balmer lines alone are less reliable in the case of
magnetic white dwarfs than in the case of non-magnetic white dwarfs,
and may be in apparent disagreement with temperature estimates derived
from the continuum slope.

We provide below effective temperature estimates derived from (a)
fitting only the Balmer lines (normalising the continuum slope of the
model spectrum to that of the observed spectrum), and (b) fitting the
continuum slope. 
% ---------- Fig. 2 -------------------------------------------------------
\begin{figure*}[t]
\includegraphics[angle=270,width=8.8cm]{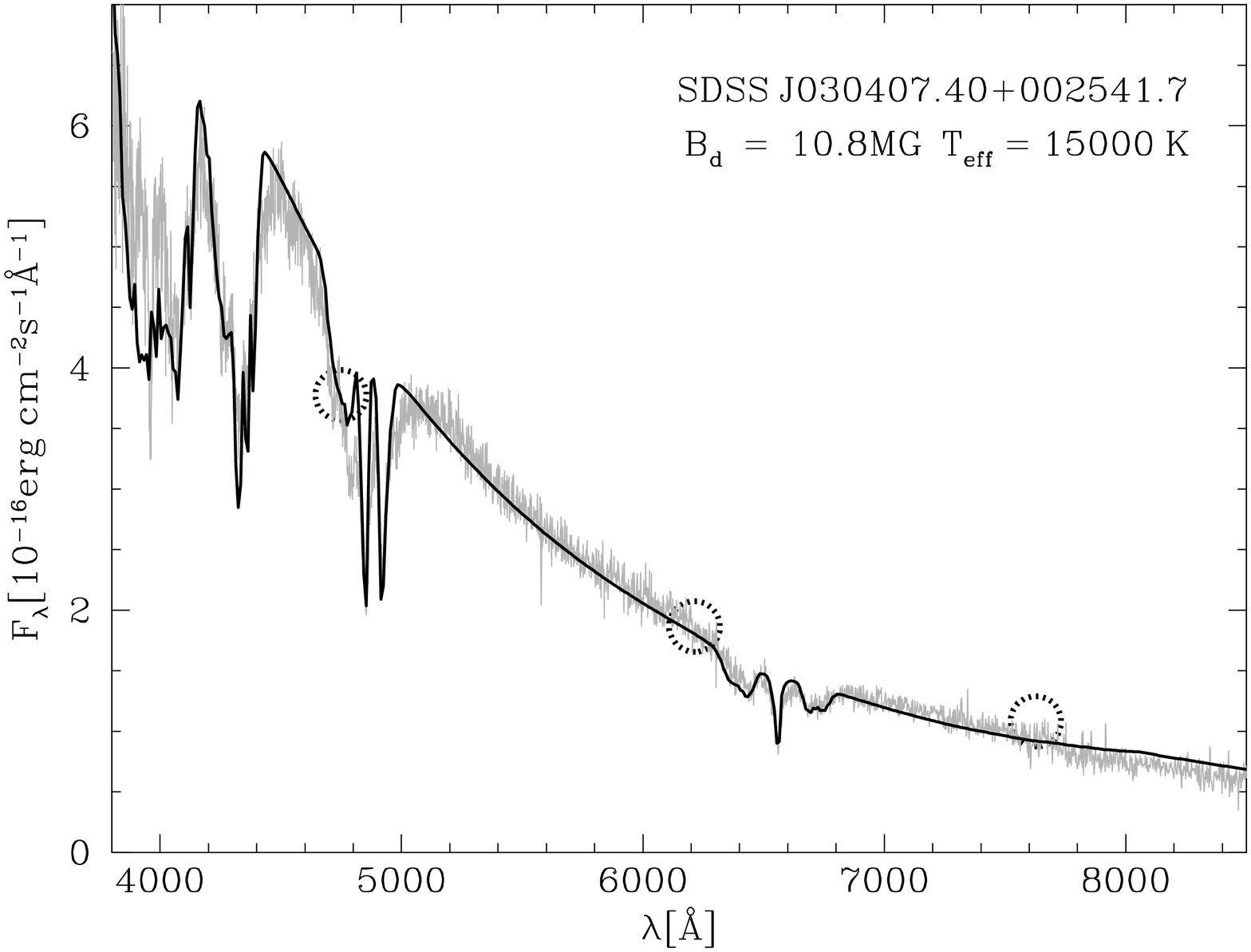}
\includegraphics[angle=270,width=8.8cm]{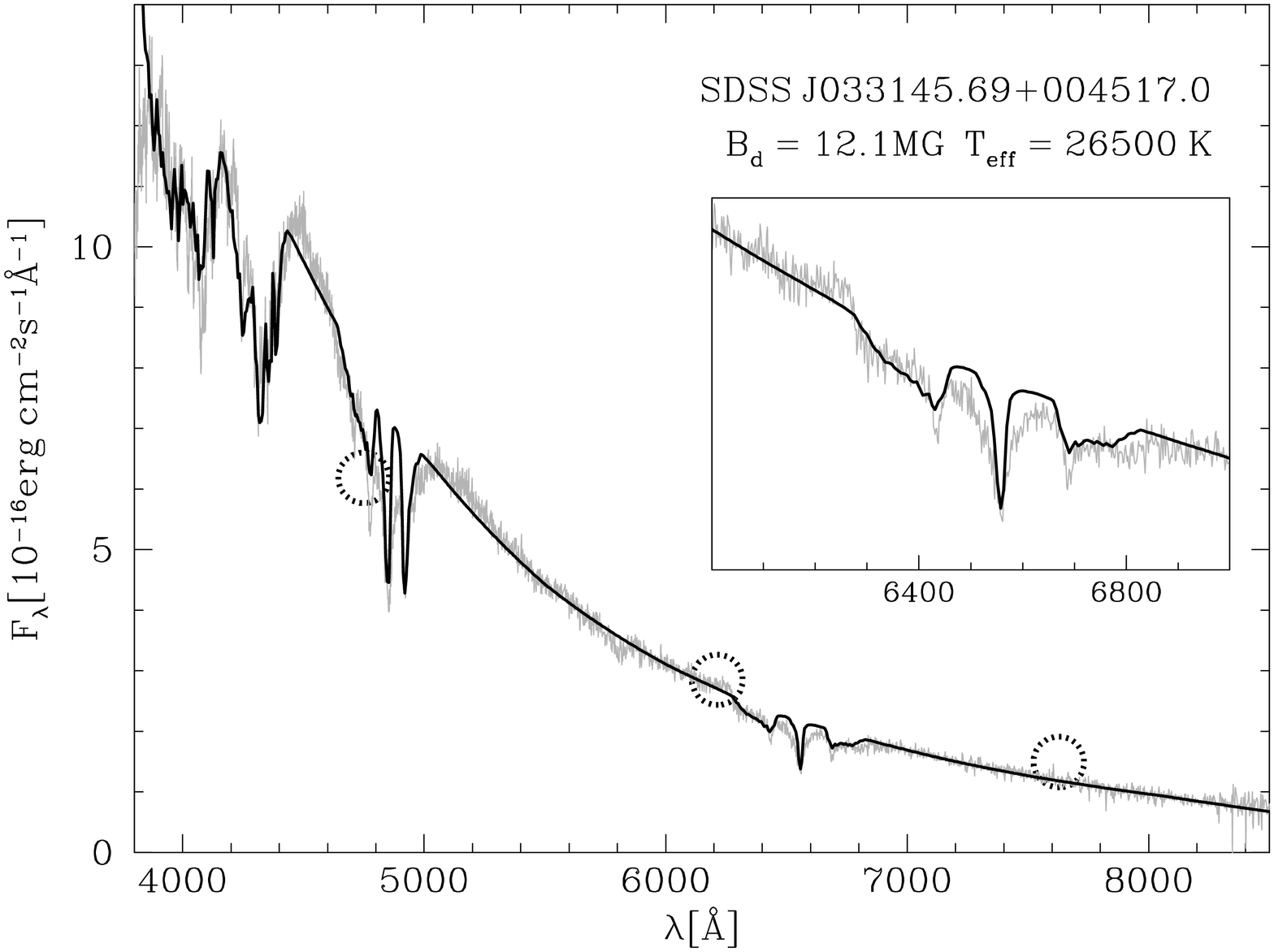}
\includegraphics[angle=270,width=8.8cm]{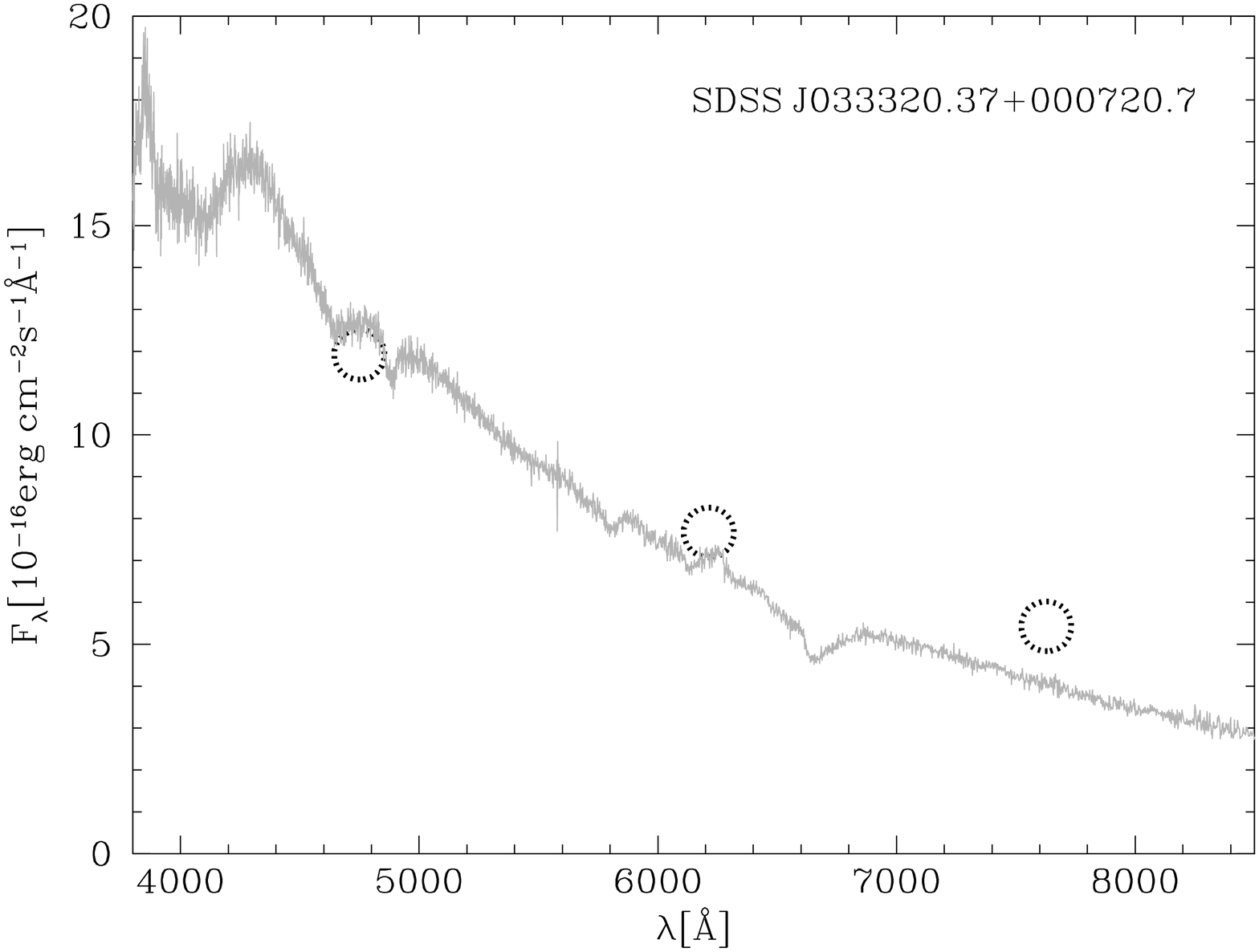}
\includegraphics[angle=270,width=8.8cm]{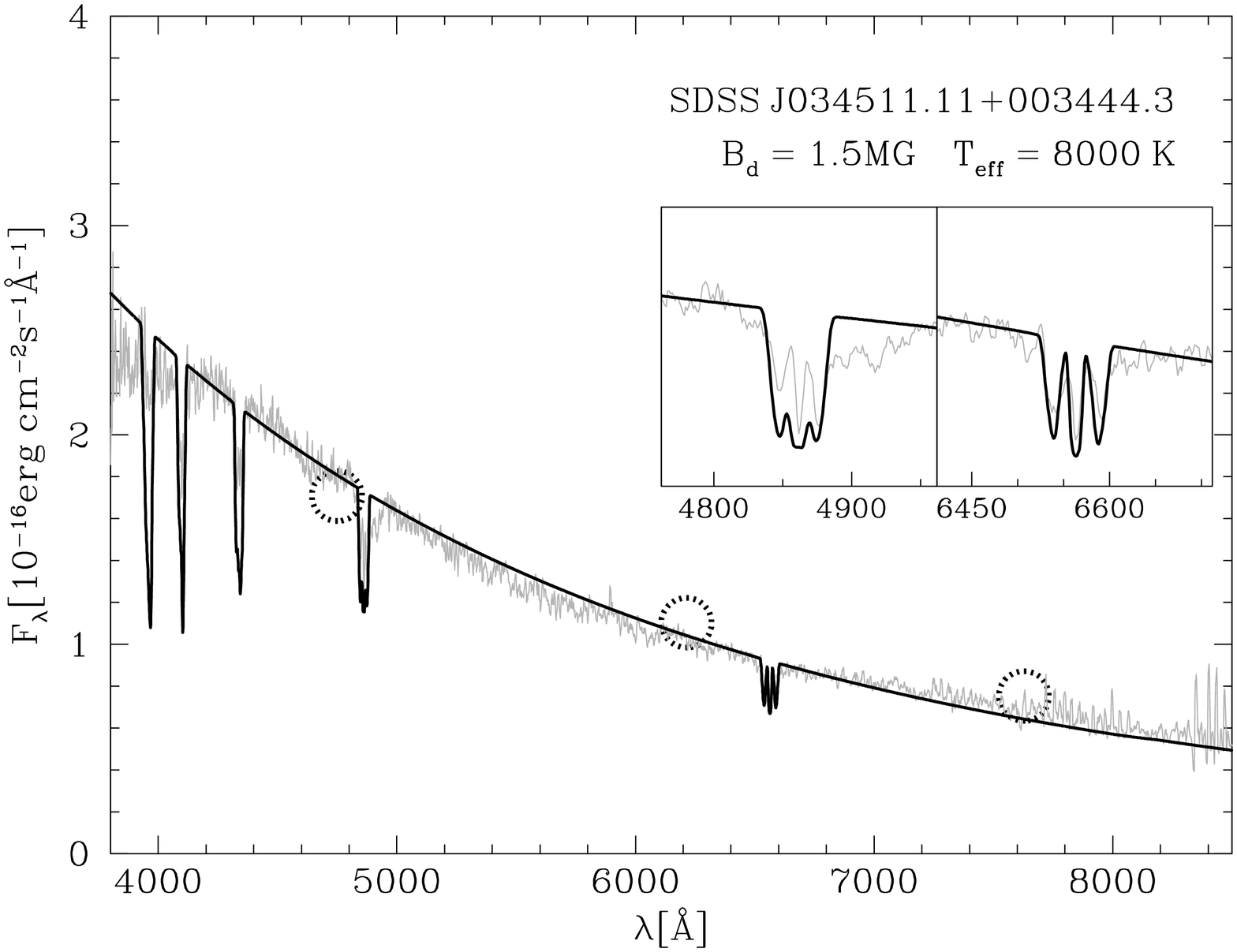}
\caption[]{\label{f-mwdspectra1} Magnetic white dwarfs from the SDSS
EDR. Gray line: SDSS fibre spectroscopy. Black line: best-fit magnetic
white dwarf model spectrum assuming a simple dipole field
configuration. Open circles: monochromatic $g^*$, $r^*$, and $i^*$
fluxes at the effective wavelengths \citep{fukugitaetal96-1}
calculated from the magnitudes of the stars in the associated imaging
data (Table\,\ref{t-sdssobs}).}
\end{figure*}
% -------------------------------------------------------------------------

% ---------- Fig. 3 -------------------------------------------------------
\begin{figure*}[t]
\includegraphics[angle=270,width=8.8cm]{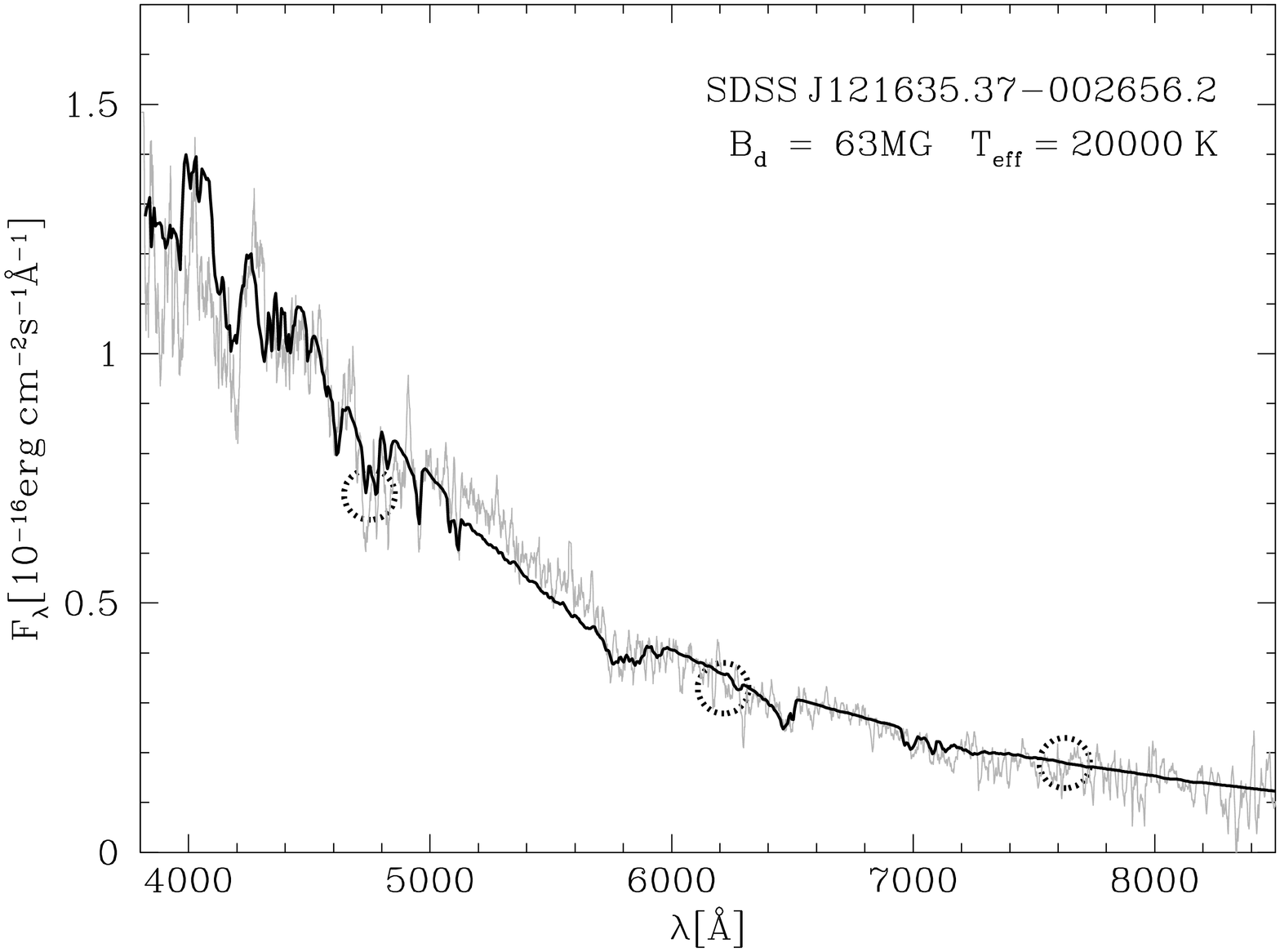}
\includegraphics[angle=270,width=8.8cm]{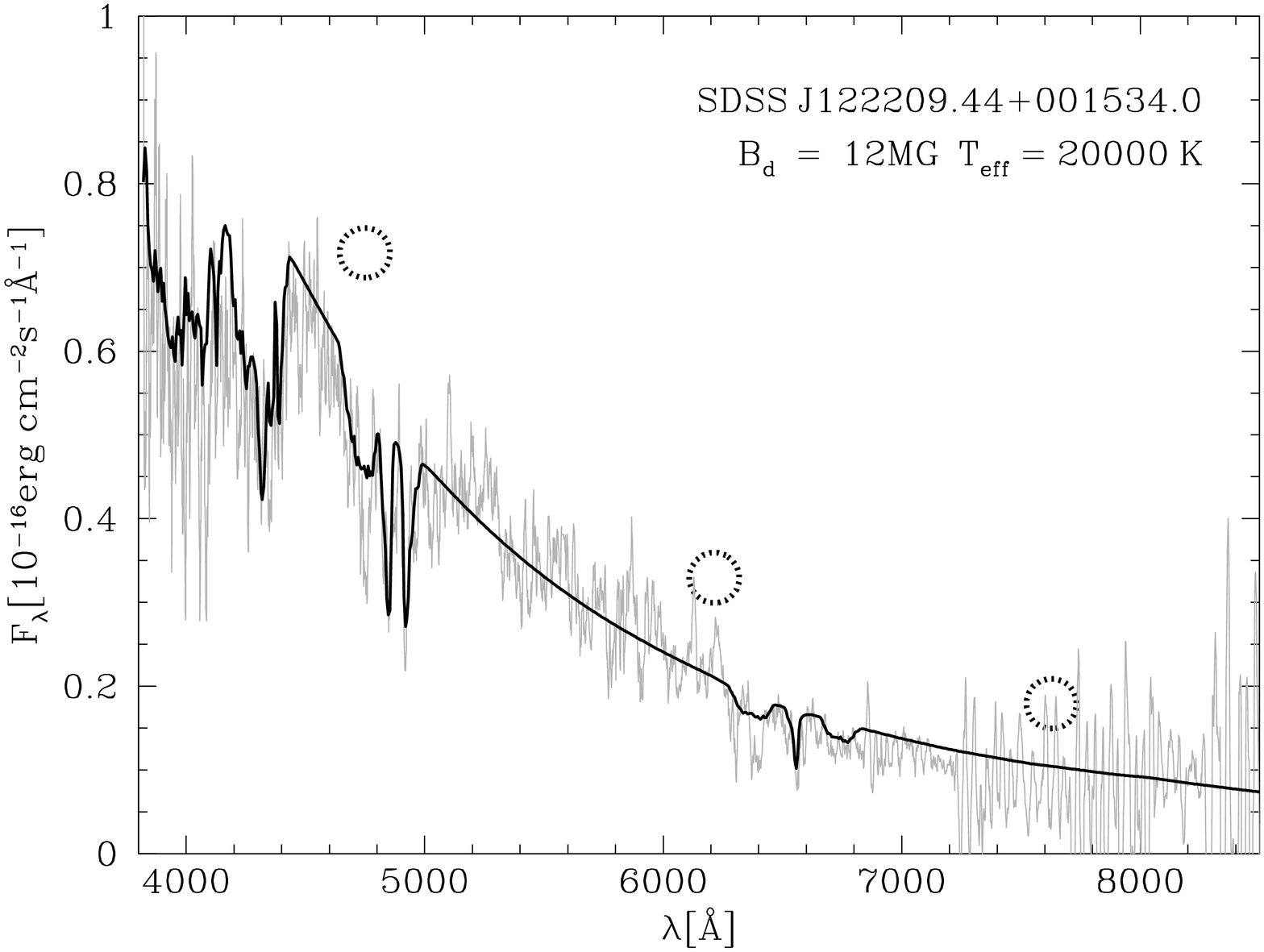}
\includegraphics[angle=270,width=8.8cm]{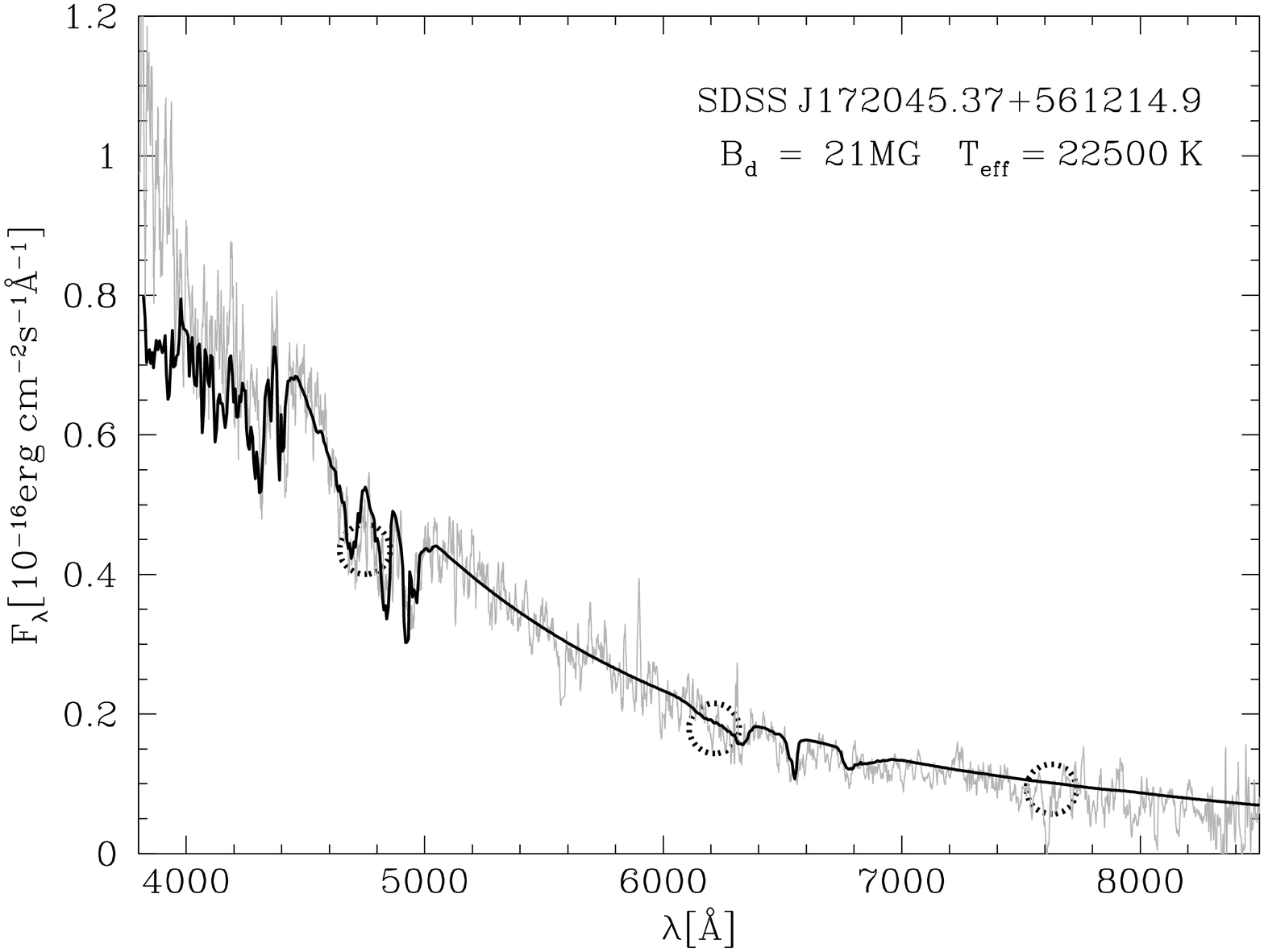}
\includegraphics[angle=270,width=8.8cm]{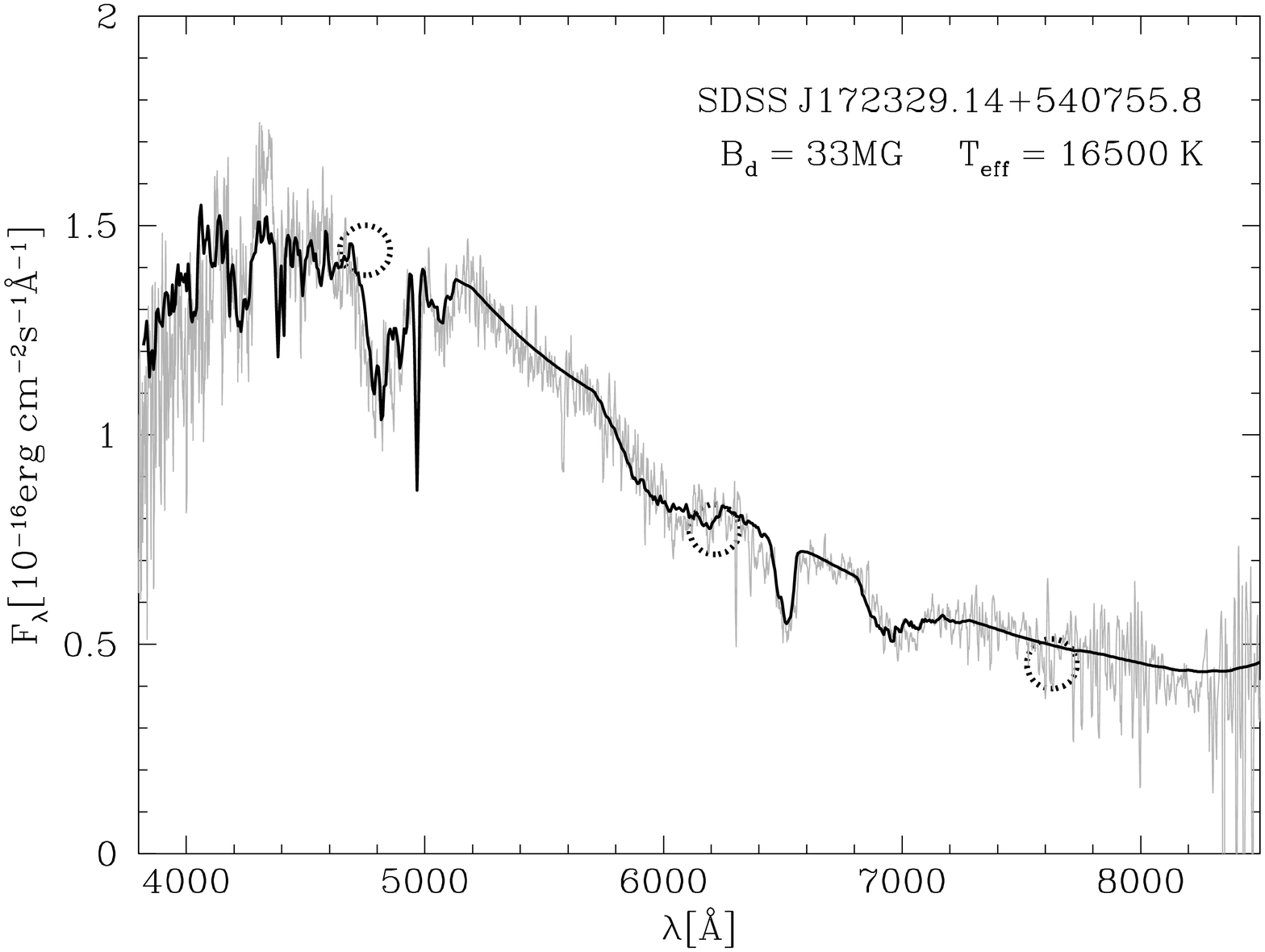}
\caption[]{\label{f-mwdspectra2} Magnetic white dwarfs from
the SDSS EDR (continued), see Fig.\,\ref{f-mwdspectra1} for further details.}
\end{figure*}
% -------------------------------------------------------------------------

\textit{SDSS\,J030407.40+002541.7:} 
The EDR spectrum of SDSS\,0304 contains the \Ha\ Zeeman triplet split
by $\sim\pm200$\,\AA, as well as numerous Zeeman components of \Hb\
and \Hg\ (Fig.\,\ref{f-mwdspectra1}). The data are well-fit with a
$\Teff=15\,000$\,K white dwarf with a dipole field strength
$B_\mathrm{d}=10.8$\,MG and a magnetic inclination
$i=50^\circ$ (where $i=90^{\circ}$ corresponds to an equator-on
view). The model spectrum provides a good match for both the continuum
slope and the strength of the Zeeman absorption lines. The most
noticeable shortcoming of this fit is the poor agreement in the
$\sigma^-$ component of \Hb. The lack of a consistent description of
the combined Zeeman and Stark effects in this magnetic field range
prevents a perfect fit to the line profiles.

\textit{SDSS\,J033145.69+004517.0\,=\,KUV\,03292+0035:} The spectrum
of SDSS\,0331 is remarkably similar to that of SDSS\,0304, and
we find from a fit to the Balmer lines $\Teff=26\,500$\,K,
$B=12.1$\,MG, and $i=55^\circ$. A fit to the continuum slope suggests
a similar temperature, $\Teff\sim25-30\,000$\,K.  \citet{jordan93-1}
derived for KUV\,03292+0035 $B_\mathrm{d}=12$\,MG, $\Teff=19\,000$\,K,
and a magnetic inclination of $i=20^{\circ}$, which is in general
agreement with the Balmer line analysis presented here. The difference
in $i$ between the two analyses might be suggestive of a rotational
variation of the Zeeman pattern, but the presently available data are
not sufficient to convincingly confirm this possibility.

Similar to SDSS\,0304, our fit to the Zeeman absorption structures
fails to reproduce the strength of the \Hb\ $\sigma^-$
component. Close inspection of the EDR spectrum of SDSS\,0331 reveals
in addition sharp $\sigma^{+/-}$ components of \Ha\ at 6435\,\AA\ and
6691\,\AA, respectively (see inset in
Fig.\,\ref{f-mwdspectra1}). These narrow features cannot be reproduced
by our simple dipole model, suggesting a more complex field
topology. Their position corresponds to a field of $\sim5.5$\,MG,
substantially lower than the dipole field strength derived from our
fit. A more detailed investigation of the field topology has to await
additional data and would be especially promising if a rotational
variation of the Zeeman features will be detected by future
measurements.

\textit{SDSS\,J033320.37+000720.7\,=\,HE\,0330--0002:} SDSS\,0333 is
one of the four magnetic white dwarfs that we have selected among the
objects classified as ``unknown'' by the SDSS processing pipeline. It
has been identified before as a peculiar white dwarf in the
Hamburg/ESO survey for bright quasars
\citep[HE\,0330--0002,][]{reimersetal98-1}. \citet{reimersetal98-1}
noted that the absorption features observed in HE\,0330--0002 are not
readily explained with either hydrogen or helium
transitions. \citet{schmidtetal01-2} confirmed HE\,0330--0002 as a
magnetic white dwarf by the detection of circular polarisation
and suggested a helium-dominated atmosphere for this star. Considering
the previous analyses of this object, we did not attempt to fit the
observed spectrum with our pure hydrogen atmosphere models.

It is interesting to note that HE\,0330--0002 appeared to be a rather
cool white dwarf in the observations of both \citet{reimersetal98-1}
and \citet{schmidtetal01-2}. The latter authors suggested
$\Teff\sim6000-7000$\,K based on the energy distribution of the
continuum. While the magnitudes derived from the SDSS imaging data are
consistent with such a low temperature, the SDSS fibre spectrum
displays a significantly bluer slope, corresponding to
$\Teff\sim12000$\,K. Considering that the SDSS imaging data was
obtained on 1998 September 25, whereas the SDSS spectroscopy was
obtained on 2000 September 23, the discrepancy in flux may suggest
intrinsic variability. We also note that the USNO-A2.0 catalogue lists
HE\,0330--0002 with $B=16.2$, whereas \citet{reimersetal98-1} give
$B_J=16.8$.  \citet{schmidtetal01-2} observed HE\,0330--0002 twice
over a few days and reported no noticeable change on that time
scale. Photometric and/or spectroscopic monitoring HE\,0330--0002 is
encouraged to test for long-term variability.

\textit{SDSS\,J034511.11+003444.3:} SDSS\,0345 is the star with the
lowest field in our sample, we find $B\simeq1.5$\,MG and
$i\simeq0^{\circ}$ from a fit to the \Ha\ and \Hb\ Zeeman
components. It is also the coldest star in our sample, the strength of
the Balmer lines suggests $\Teff\simeq8000$\,K, whereas a fit to the
continuum results in $\Teff\simeq9000$\,K.

\textit{SDSS\,J121635.37--002656.2:}
SDSS\,1216 has been selected as a magnetic white dwarf candidate among
the objects classified as ``unknown'' in the SDSS EDR database because
of the complex absorption structure near \Hb\ and \Hg\ seen in its
spectrum. A dipole fit to the Balmer lines confirms the object as a
magnetic white dwarf with $B_\mathrm{d}=63$\,MG, the highest field in
the sample of magnetic white dwarfs presented here,
$\Teff=20\,000$\,K, and $i=75^{\circ}$. The effective temperature
derived from the fit to the Balmer lines is somewhat higher than that
derived from the continuum slope, $\Teff=15\,000$\,K.

\textit{SDSS\,J122209.44+001534.0:} 
SDSS\,1222, classified as ``unknown'' in the SDSS EDR database, is the
faintest object in our sample of SDSS magnetic white dwarfs and the
signal-to-noise ratio of the  fibre spectrum is consequently very
low. Nevertheless, the Zeeman triplet of \Hb\ split in a moderate
field is clearly recognised, and we find $B=12$\,MG,
$\Teff=20\,000$\,K, and $i=40^\circ$.  The effective temperature
derived from the fit to the Balmer lines is broadly consistent with
the continuum slope.

\textit{SDSS\,J172045.37+561214.9:}
The spectrum of SDSS\,1720, the fourth magnetic white dwarf candidate
selected from the ``unknown'' class in the SDSS EDR database, displays
a broad \Hb\ triplet. The parameters derived from a dipole fit to the
Balmer lines are are $B_\mathrm{d}=21$\,MG, $\Teff=22500$\,K, and
$i=85^\circ$. The slope of the continuum is consistent with a
temperature in the range 20--30\,000\,K.
 
\textit{SDSS\,J172329.14+540755.8:} SDSS\,1723 has been selected
because of the broad \Ha\ triplet and the complex \Hb\ absorption
detected in its spectrum. The turnover of the continuum at
$\lambda\la4200$\,\AA\ indicates a relatively low temperature. Our
dipole fit to the Balmer lines results in $B_\mathrm{d}=33$\,MG,
$\Teff=16\,500$\,K, and $i=35^\circ$. Fitting the continuum slope of the
spectrum of SDSS\,1723 results in a significantly lower value for the
effective temperature, $\Teff\simeq10\,000$\,K.

\textit{SDSS\,J232248.22+003900.9:}
The fibre spectrum of SDSS\,2322 contains relatively weak but broad
absorption structures near \Hb\ and \Hg\ in a blue
continuum. A dipole fit confirms the object as a magnetic white dwarf
with $B_\mathrm{d}=13$\,MG, $\Teff=39\,000$\,K, and $i=25^\circ$. The
temperature derived from the continuum slope is
$\Teff\simeq25-30\,000$\,K.

\begin{figure*}[t]
\includegraphics[angle=270,width=8.8cm]{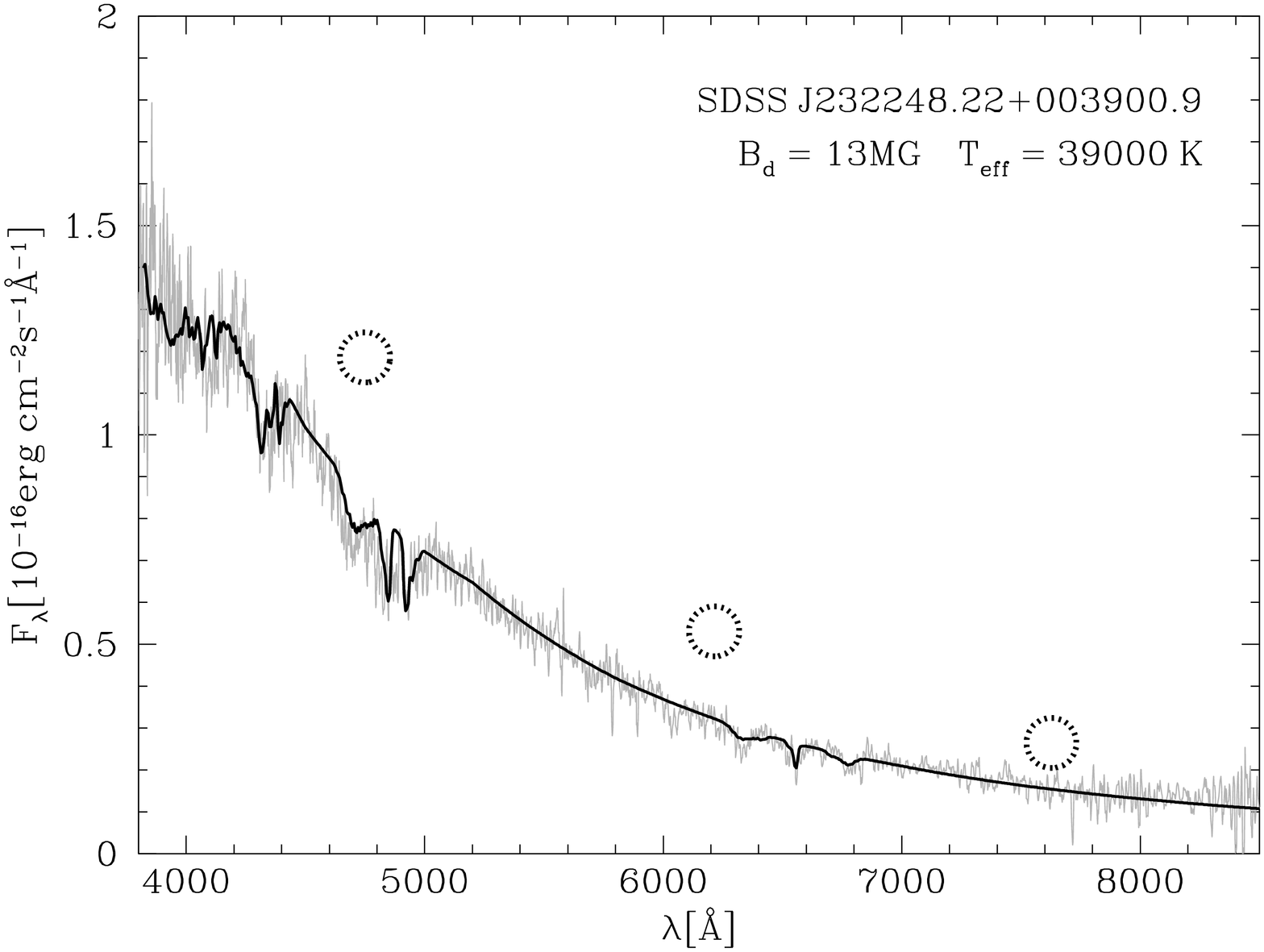}
\includegraphics[angle=270,width=8.8cm]{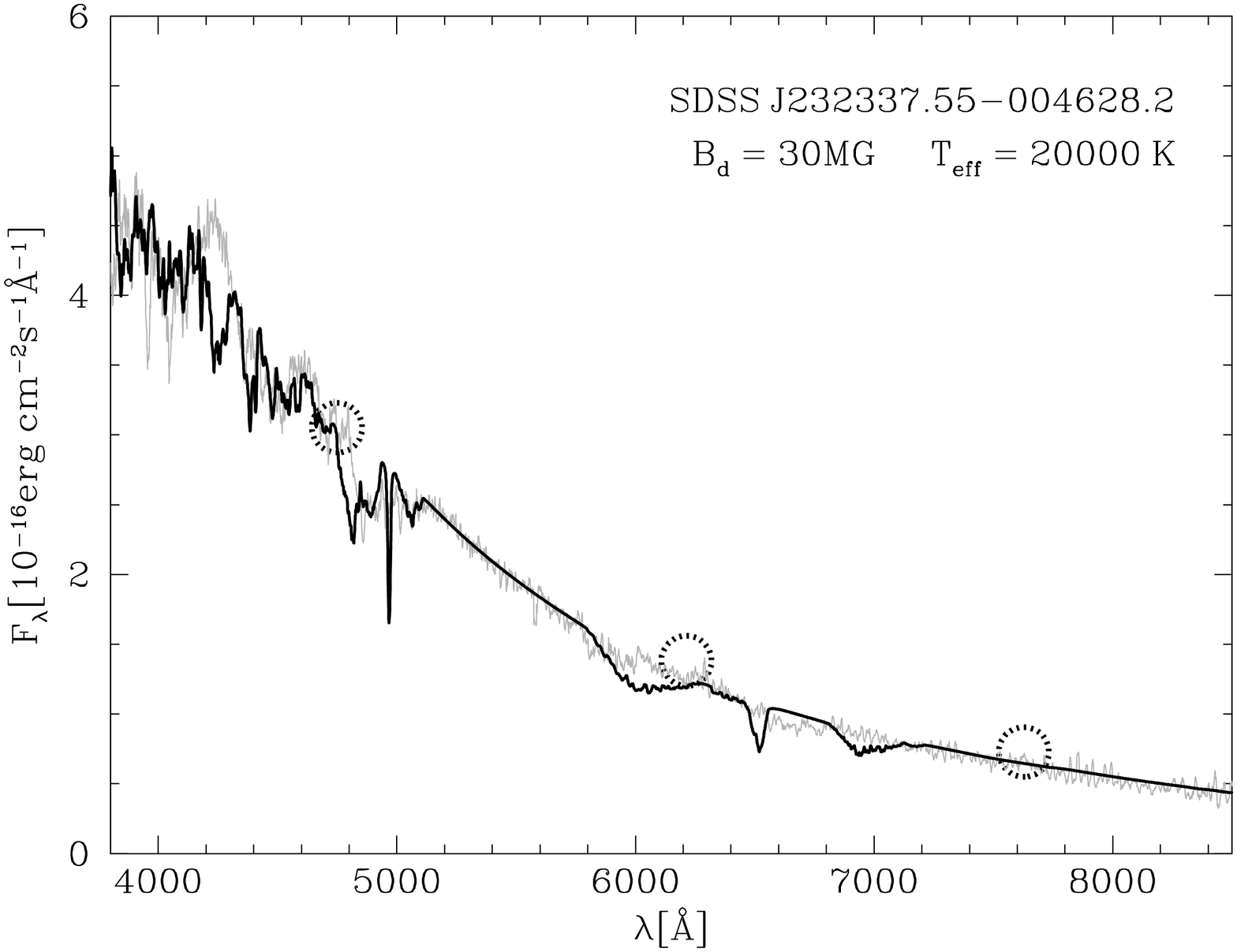}
\caption[]{\label{f-mwdspectra3} Magnetic white dwarfs from
the SDSS EDR (continued), see Fig.\,\ref{f-mwdspectra1} for further details.}
\end{figure*}

\textit{SDSS\,J232337.55--004628.2:} 
The spectrum of SDSS\,2323 resembles that of SDSS\,1222 with broad
absorption features of \Hb/\Hg\ and a blue continuum. However, we were
not able to find a satisfying fit with our dipole
models. Figure\,\ref{f-mwdspectra3} shows the fibre spectrum of
SDSS\,2323 along with an illustrative dipole model computed for
$B=30$\,MG, $\Teff=20\,000$\,K, and $i=0^\circ$.  Close
inspection of the spectrum reveals that a number of the narrow
absorption lines present in the spectrum could be explained with He\,I
transitions. However, while the width of the \Hb\ and \Hg\ absorption
troughs clearly suggests the (magnetic?) white dwarf nature of this
object, the potential He\,I lines are too narrow for an origin in a
high gravity atmosphere. Inspection of the SDSS $r^*$ image did not
reveal anything suspicious about this object (e.g. a close visual
companion). Additional observations of SDSS\,2323 are needed to
clarify the nature of this object.

% ---------- Fig. 4 -------------------------------------------------------
% -------------------------------------------------------------------------

% ---------- Table 2 -------------------------------------------------------
\begin{table}[t]
\caption[]{\label{t-damwd} Best-fit parameter for the confirmed DA
magnetic white dwarfs. Two effective temperature estimates are given,
based in fits to the Balmer lines and to the continuum slope.}
\begin{flushleft}
\begin{tabular}{lrrr}
\hline\noalign{\smallskip}
MWD  & $B_\mathrm{d}$\,[MG] & \Teff\,[K] & \Teff\,[K]  \\ 
     &                      & lines      & continuum   \\
\hline\noalign{\smallskip}
SDSS\,J030407.40--002541.7   & 10.8 & 15\,000 & 15\,000 \\ 
SDSS\,J033145.69+004517.0    & 12.1 & 26\,500 & 25-30\,000   \\
SDSS\,J034511.11+003444.3    & 1.5  & 8000    & 9000 \\
SDSS\,J121635.37--002656.2   & 63   & 20\,000 & 15\,000 \\
SDSS\,J122209.44+001534.0    & 12   & 20\,000 & 15-20\,000 \\
SDSS\,J172045.37+561214.9    & 21   & 22\,500 & 20-30\,000\\
SDSS\,J172329.14+540755.8    & 33   & 16\,500 & 10\,000 \\ 
SDSS\,J232248.22+003900.9    & 13   & 39\,000 & 25-30\,000\\
\noalign{\smallskip}\hline
\end{tabular}
\end{flushleft}
\end{table}
% -------------------------------------------------------------------------

\section{Discussion \& Conclusions}
The population of known magnetic white dwarfs presently comprises
$\sim65$ stars \citep{wickramasinghe+ferrario00-1}. Considering that
the first magnetic white dwarf (Grw$+70^{\circ}8047$) has been
discovered nearly 60 years ago, the ``average'' discovery rate of
these stars has been $\sim1$ per year. As outlined in the
Introduction, this sample is still far too small to stringently test
theoretical models for the origin of the magnetic field and for the
evolution of magnetic white dwarfs.

Unfortunately, discovering new magnetic white dwarfs is a tedious
process. Figure\,\ref{f-colours} shows colour-colour diagrams for all
``stellar'' objects from the EDR spectroscopic database. It is
apparent that the magnetic white dwarfs discussed are well distributed
over the locus of hot stars --~mainly white dwarfs and subdwarfs. It
is, hence, not possible to select magnetic white dwarfs by their
colour alone, the availability of a large spectroscopic data set is
essential for a significant increase in the number of known magnetic
white dwarfs.

In this paper, we have presented fibre spectroscopy of 10 magnetic
white dwarf candidates identified in the EDR of the SDSS. We confirm
eight of these objects as magnetic DA white dwarfs, of which seven are
new discoveries spanning a wide range of magnetic field
strengths. This represents a substantial addition to the population of
known magnetic white dwarfs. The spectrum of SDSS\,2323 contains
relatively broad \Hb\ and \Hg\ absorption troughs, however, we were
not able to confirm the magnetic nature of this star from the SDSS
data alone. Finally, we recovered the peculiar magnetic white dwarf
HE\,0330--0002.

Our eye ball inspection of the SDSS fibre spectra is certainly biased
in a number of ways.  In white dwarfs with weak magnetic field
strengths ($B\la1-2$\,MG) the resolution and the signal-to-noise (S/N)
ratio of the SDSS spectra becomes insufficient to detect the Zeeman
splitting of the Balmer lines.  Insufficient S/N of the SDSS also
lowers the probability of detecting the weak Zeeman absorption
components in both hot or high field white dwarfs
($\Teff\ga40\,000$\,K, $B\ga80$\,MG). Finally we have
restricted or analysis to magnetic white dwarfs with a pure hydrogen
atmosphere~--~and discarded a handful of SDSS spectra with absorption
features that could not be identified either with typical stellar
transitions or with hydrogen Zeeman components for a wide range of
field strengths. We are, however, confident that our selection of
magnetic white dwarfs is complete for stars with pure hydrogen
atmospheres, magnetic fields of a few MG to a few tens MG, and
effective temperatures $10\,000\la\Teff\la40\,000$\,K. An \textit{a
posteriori} check of the EDR database using the lists of
\citet{jordan01-1} and \citet{wickramasinghe+ferrario00-1} confirmed
that KUV\,\,03292+0035 and HE\,0330--0002 are the only previously
known magnetic white dwarfs with spectroscopic coverage.

The Early Data Release represents $\sim$5\% of the final SDSS data
set. Accounting for our selection bias, we conservatively estimate
that the complete SDSS will increase the number of known magnetic
white dwarfs by at least a factor 3. Scaling the number of white
dwarfs identified by \citet{stoughtonetal02-1} in the EDR sample (734)
to the complete SDSS, and assuming $\sim2-4\%$ as ratio of magnetic to
non-magnetic white dwarfs \citep{jordan01-1,
wickramasinghe+ferrario00-1}, suggests that a careful exploitation of
the entire SDSS data base, including intense spectroscopic and
polarimetric follow-up observations, may well lead to the discovery of
several hundred new magnetic white dwarfs.

\acknowledgements BTG was supported by a PPARC Advanced Fellowship, FE
was supported by the DLR under grant 50\,OR\,99\,03\,6, SJ was
supported by the DLR under grant 50\,OR\,0201. We thank the
referee, Pierre Bergeron, for suggesting the comparison of the
observed SDSS EDR colours with the theoretical white dwarf cooling
tracks, and for providing the synthetic  $u^*g^*r^*i^*z^*$ colours
used in the bottom panel of Fig.\,\ref{f-colours}. Susanne Friedrich
is acknowledged for providing the coordinates of most previously known
white dwarfs.

The Sloan Digital Sky Survey (SDSS) is a joint project of the
University of Chicago, Fermilab, the Institute for Advanced Study, the
Japan Participation Group, the Johns Hopkins University, the
Max-Planck-Institut f\"{u}r Astronomie, New Mexico State University,
Princeton University, the United States Naval Observatory, and the
University of Washington.  Apache Point Observatory, site of the SDSS,
is operated by the Astrophysical Research Consortium.  Funding for the
project has been provided by the Alfred P. Sloan Foundation, the SDSS
member institutions, the National Aeronautics and Space
Administration, the National Science Foundation, the U.S. Department
of Energy, Monbusho, and the Max Planck Society.  The SDSS World Wide
Web site is http://www.sdss.org/.

%\bibliographystyle{aa}
%\bibliography{aamnem99,$HOME/tex/Papers/Bibliography/aabib}

\end{document}